\documentclass{article}
\usepackage{epsfig}
\usepackage{amsmath}
\usepackage{amsfonts}\tolerance=10000
\usepackage{graphicx}
\usepackage[german, english]{babel}
\usepackage{a4wide}
\usepackage{amsmath}
\usepackage{amssymb}
\usepackage{ifthen}

\date{}

\newcommand{\la}{\lambda}
\newcommand{\ka}{\kappa}
\newcommand{\al}{\alpha}

\newcommand{\Ga}{\Gamma}

\newcommand{\si}{\sigma}
\newcommand{\Si}{\Sigma}
\newcommand{\f}{\phi}
\newcommand{\vf}{\varphi}
\newcommand{\F}{\Phi}

\newcommand{\bnabla}{\mbox{\boldmath $\nabla$}}

\newcommand{\bOmega}{\mbox{\boldmath $\Omega$}}

\newcommand{\ee}{\end{equation}}
\newcommand{\eea}{\end{eqnarray}}
\newcommand{\be}{\begin{equation}}
\newcommand{\bea}{\begin{eqnarray}}

\newcommand{\pa}{\partial}

\newcommand{\vep}{\varepsilon}

\newcommand{\re}[1]{(\ref{#1})}
\newcommand{\R}{{\rm I \hspace{-0.52ex} R}}

\newcommand{\eins}{1\hspace{-0.56ex}{\rm I}}

\numberwithin{equation}{section}
\begin{document} 

\title{New Chern-Simons densities in both odd and even dimensions} 
 
\author{
{\large Eugen Radu}$^{\dagger \star}$ and
{\large Tigran Tchrakian}$^{\ddagger \star \diamond}$ \\ \\
\\
$^{\ddagger}${\small  School of Theoretical Physics -- DIAS, 10 Burlington
Road, Dublin 4, Ireland}
\\
$^{\star}${\small Department of Computer Science,
National University of Ireland Maynooth,
Maynooth,
Ireland}\\
$^{\diamond}${\small Theory Division, Yerevan Physics Institute (YerPhI),
AM-375 036 Yerevan 36, Armenia} }
 
\maketitle 
\begin{abstract}
After reviewing briefly the dimensional reduction of Chern--Pontryagin densitie, we define new Chern--Simons
densities expressed in terms of Yang-Mills and Higgs fields.
These are defined in all dimensions, including in even dimensional
spacetimes. They are constructed by subjecting the dimensionally reduced
Chern--Pontryagin densites to further descent by two steps.
\end{abstract}

\section{Introduction}

The central task of these notes is to explain how to subject $n-$th Chern--Pontryagin (CP)
density ${\cal C}^{(n)}$,
\be
\label{CP}
{\cal C}^{(n)}=\frac{1}{\omega(\pi)}\vep_{M_1M_2M_3M_4
\dots M_{2n-1}M_{2n}}\mbox{Tr}\,F_{M_1M_2}F_{M_3M_4}\dots F_{M_{2n-1}M_{2n}} 
\ee
defined on on the $2n-$dimensional space, to dimensional descent to $\R^D$ by considering \re{CP} on
the direct product space $\R^D\times S^{2n-D}$. The resulting residual density on $\R^D$ will be denoted as
$\Omega_{D}^{(n)}$.

The density ${\cal C}^{(n)}$ is by construction, a total divergence
\be
\label{totdivn}
{\cal C}^{(n)}=\bnabla\cdot\bOmega^{(n)}\,,
\ee
and it turns out that under certain retrictions, the dimensional descendant of \re{totdivn} is also a total divergence.
This result will be applied to the construction of Chern-Simons solitons
in all dimensions.

Some special choices, or restrictions, are made for practical reasons.
Firstly, we have restricted to the codimension $S^{2n-D}$, the $(2n-D)-$sphere, as this is the most symmetric
compact coset space that is defined both in even and in odd dimensions. It can of course be replaced by any other
symmetric and compact coset space.

Secondly, the gauge field of the bulk gauge theory is chosen to be a $2^{n-1}\times 2^{n-1}$ array with complex valued
entries. Given our choice of spheres for the codimension, this leads to residual gauge fields on which take their
values in the Dirac matrix representation of the residual gauge group $SO(D)$.

As a result of the above two choices, it is possible to make the symmetry imposition (namely the dimensional reduction)
such that the residual Higgs field is described by a $D-$component isovector
multiplet. With this choice, the asymptotic gauge
fields describe a Dirac-Yang~\cite{D,yang,Tchrakian:2008zz} monopoles..

The central result exploited here is that the CP density ${\cal C}_{D}^{(n)}$
on $\R^D$ descended from the $2n-$dimensional bulk CP density ${\cal C}^{(n)}$ is {\bf also} a $total$ $divergence$
\be
\label{totdiv}
{\cal C}_{D}^{(n)}=\bnabla\cdot\bOmega^{(n,D)}
\ee
like ${\cal C}^{(n)}$ formally is on the bulk.
From the reduced density $\bOmega^{(n,D)}\equiv\Omega^{(n,D)}_i$, where the index $i$ labels the coordinate of the
residual space $x_i$ with $i=1,2,\dots,D$, one can identify a Chern--Simons
(CS) density as the $D-$th component of $\bOmega^{(n,D)}$. This quantity can then be 
interpreted as a CS term on $(D-1)-$ dimensional Monkowski space, $i.e.$,
on the spacetime $(t,\R^{D-2})$. The solitons of the
corresponding CS--Higgs (CSH) theory can be constructed systematically.
Note, that this is not the usual CS term defined in terms of a pure Yang--Mills field on
$odd$ dimensional spacetime, but rather these new CS terms are defined by both the YM and,
the Higgs fields. Most importantly, the definition of these new CS terms is not restricted to
$odd$ dimensional spacetimes, but covers also $even$ dimensional spacetimes. 
Such CSH solutions have not been studied
to date.

\section{Dimensional reduction of gauge fields}
\label{YMdescent}
The calculus of the dimensional reduction of Yang--Mills fields employed here is based on the formalism of
A.~S.~Schwarz~\cite{Schwarz:1977ix},
which is specially transparent due to the choice of displaying the results only at a fixed point of the compact
symmetric codimensional space $K^N$ (the North or South pole for $S^N$). Our formalism is a straightforward extension
of \cite{Schwarz:1977ix,Romanov:1977rr,Schwarz:1981mb}.

\subsection{Descent over $S^N$: $N$ odd}
\label{YMdescentodd}
For the descent from the bulk dimension $2n=D+N$ down to {\bf odd} $D$ (over odd $N$),
the components of the residual connection evaluated at the Noth pole of $S^N$ are given by
\bea
{\cal A}_i&=&A_i(\vec x)\otimes\eins\label{aiodd}\\
{\cal A}_I&=&\F(\vec x)\otimes\frac12\Gamma_I\,.\label{aIodd}
\eea
The unit matrix in \re{aiodd}, like the $N-$dimensional gamma matrix in \re{aIodd}, are
$2^{\frac12(N-1)}\times 2^{\frac12(N-1)}$ arrays. Choosing the $2^{n-1}\times 2^{n-1}$ bulk gauge group to be, say,
$SU(n-1)$, allows the choice of $SO(D)$ as the gauge group of the residual connection $A_i(x)$. This choice is
made such that the asymptotic connections describe a Dirac--Yang monopole.

For the same reason, the choice for the multiplet structure of the Higgs field is made to be less restrictive.
The (anti-Hermitian) field $\F$, which is not necessarily traceless~\footnote{In practice, when constructing soliton
solutions, $\F$ is taken to be traceless
without loss of generality.}, can be and $is$
taken to be in the algebra of $SO(D+1)$, in particular, in one or other of the chiral
reprentations of $SO(D+1)$, $D+1$ here being even.
\be
\label{PHYodd}
\F=\f^{ab}\,\Si_{ab}\ ,\quad a=i,\,D+1\ ,\quad i=1,2,\dots, D\,.
\ee
(Only in the $D=3$ case does the Higgs field take its values
in the algebra of $SO(3)$, since the representations $SO(3)$ coincide with those of chiral $SO(4)$.)

In anticipation of the corresponding situation of even $D$ to be given next, one can specialise \re{PHYodd}
to the a $D-$component $isovector$ expression of the Higgs field
\be
\label{Hisoodd}
\F=\f^i\,\Si_{i,D+1}\,,
\ee
with the purpose of having a unified notation for both even and odd $D$, where the Higgs field takes its values
in the components $\Si_{i,D+1}$ orthogonal to elements $\Si_{ij}$ of the algebra of $SO(D+1)$.
This specialisation is not necessary, and
is in fact inappropriate should one consider, $e.g.$, axially symmetric fields. It is however adequate for the
presentation here    and is sufficiently general to describe spherically symmetric
monopoles~\footnote{While all concrete considerations in the following are restricted to spherically symmetric fields,
it should be emphasised that relaxing spherical symmetry results in the Higgs multiplet getting out of the orthogonal
complement $\Si_{i,D+1}$ to $\Si_{i,j}$. Indeed, subject to axial symmetry one has
\be
\label{ax}
\F=f_1(\rho,z)\Si_{\al\beta}\hat x_{\beta}+f_2(\rho,z)\Si_{\beta,D+1}\hat x_{\beta}+f_3(\rho,z)\Si_{D,D+1}\,,
\ee
where $x_i=(x_{\al},z)$, $|x_{\al}|^2=\rho^2$ and with $\hat x_{\al}=x_{\al}/\rho$. Clearly, the term in \re{ax}
multiplying the basis $\Si_{\al\beta}$ does not occur in \re{Hisoodd}.}.

In \re{aiodd} and \re{aIodd}, and everywhere henceforth,
we have denoted the components of the residual coordinates as $x_i=\vec x$.
The dependence on the codimension coordinate $x_I$ is suppressed since all fields are evaluated at a fixed point
(North or South pole) of the codimension space.

The resulting components of the curvature are
\bea
{\cal F}_{ij}&=&F_{ij}(\vec x)\otimes\eins\label{fijodd}\\
{\cal F}_{iI}&=&D_{i}\F(\vec x)\otimes\frac12\Gamma_I\label{fiIodd}\\
{\cal F}_{IJ}&=&S(\vec x)\,\otimes\Gamma_{IJ}\,,\label{fIJodd}
\eea
where $\Gamma_{IJ}=-\frac14[\Gamma_{I},\Gamma_{J}]$ are the Dirac representation matrices of $SO(N)$, the stability
group of the symmetry group of the $N-$sphere. In \re{fiIodd}, $D_{i}\F$ is the covariant derivative of the
Higgs field $\F$
\be
\label{covodd}
D_{i}\F=\pa_i\F+[A_i,\F]
\ee
and $S$ is the quantity
\be
\label{Sodd}
S=-(\eta^2\,\eins+\F^2)\,,
\ee
where $\eta$ is the inverse of the radius of the $N-$sphere.

\subsection{Descent over $S^N$:  $N$ even}
\label{YMdescenteven}
The formulae corresponding to \re{aiodd}-\re{fIJodd} for the case of {\bf even} $D$ are somewhat more complex.
The reason is the existence of a chiral matrix $\Gamma_{N+1}$, in addition to the Dirac matrices $\Gamma_{I}$,
$I=1,2,\dots,N$. Instead of \re{aiodd}-\re{aIodd} we now have
\bea
{\cal A}_i&=&A_i(\vec x)\otimes\eins+B_i(\vec x)\otimes\Gamma_{N+1}\nonumber\\
{\cal A}_I&=&\f(\vec x)\otimes\frac12\Gamma_I+\psi(\vec x)\otimes\frac12\Gamma_{N+1}\Gamma_I\nonumber\,,
\eea
where $A_i$, $B_i$, $\f$, and $\psi$ are again antihermitian matrices, but with only $A_i$ being traceless.
The fact that $B_i$ is not traceless here results in an Abelian gauge field in the reduced system.

Anticipating what follows, it is much more transparent to re-express these formulas in the form
\bea
{\cal A}_i&=&A_i^{(+)}(\vec x)\otimes\,P_++A_i^{(-)}(\vec x)\otimes\,P_-
+\frac{i}{2}\,a_i(\vec x)\,\Gamma_{N+1}\label{aieven}\\
{\cal A}_I&=&\vf(\vec x)\otimes\frac12\,P_+\,\Gamma_I-\vf(\vec x)^{\dagger}\otimes\frac12\,P_-\,\Gamma_I\label{aIeven}\,,
\eea
where now $P_{\pm}$ are the $2^{\frac{N}{2}}\times 2^{\frac{N}{2}}$ projection operators
\be
\label{proj}
P_{\pm}=\frac12\left(\eins\pm\Gamma_{N+1}\right)\,.
\ee
In \re{aieven}, the residual gauge connections $A_i^{(\pm)}$ are anti-Hermitian and traceless
$2^{\frac{D}{2}}\times 2^{\frac{D}{2}}$ arrays, and the Abelian connection $a_i$ results directly from the trace of the
field $B_i$. The $2^{\frac{D}{2}}\times 2^{\frac{D}{2}}$ "Higgs'' field $\vf$ in \re{aIeven} is neither Hermitian nor
anti-Hermitian. Again, to achieve the desired breaking of the gauge group, to lead eventually to the requisite Higgs
$isomultiplet$, we choose the gauge group in the bulk to be $SU(n-1)$, where $2n=D+N$.

The components of the curvaturs are readily calculated to give
\bea
{\cal F}_{ij}&=&F_{ij}^{(+)}(\vec x)\otimes\,P_++F_{ij}^{(-)}(\vec x)\otimes\,P_-
+\frac{i}{2}\ f_{ij}(\vec x)\,\Gamma_{N+1}\label{fijeven}\\
{\cal F}_{iI}&=&D_i\vf(\vec x)\otimes\,\frac12\,P_+\Gamma_I
-D_i\vf^{\dagger}(\vec x)\otimes\,\frac12\,P_-\Gamma_I\label{fiIeven}\\
{\cal F}_{IJ}&=&S^{(+)}(\vec x)\otimes\,P_+\Gamma_{IJ}+S^{(-)}(\vec x)\otimes\,P_-\Gamma_{IJ}\,,\label{fIJeven}
\eea
the curvatures in \re{fijeven} being defined by
\bea
F_{ij}^{(\pm)}&=&\pa_iA_j^{(\pm)}-\pa_jA_i^{(\pm)}+[A_i^{(\pm)},A_j^{(\pm)}]\label{curvpm}\\
f_{ij}&=&\pa_ia_j-\pa_ja_i\,.\label{curvabel}
\eea
The covariant derivative in \re{fiIeven} now is defined as
\bea
D_i\vf&=&\pa_i\vf+A_i^{(+)}\,\vf-\vf\,A_i^{(-)}+i\,a_i\,\vf\label{coveven}\\
D_i\vf^{\dagger}&=&\pa_i\vf^{\dagger}+A_i^{(-)}\,\vf^{\dagger}-\vf^{\dagger}\,A_i^{(+)}
-i\,a_i\,\vf^{\dagger}\,,\label{covevendag}
\eea
and the quantities $S^{(\pm)}$ in \re{fIJeven} are
\be
\label{Spm}
S^{(+)}=\vf\,\vf^{\dagger}-\eta^2\quad,\quad S^{(-)}=\vf^{\dagger}\,\vf-\eta^2\,.
\ee

In what follows, we will suppress the Abelian field $a_i$, since only when less stringent symmetry than spherical is
imposed is it that it would contribute. In any case, using the formal replacement
\[
A_i^{(\pm)}\leftrightarrow A_i^{(\pm)}\pm\frac{i}{2}\,a_i\,\eins
\]
yields the algebraic results to be derived below, in the general case.

We now refine our calculus of descent over even codimensions further. We see from \re{aieven} that $A_i^{(\pm)}$ being
$2^{\frac{D}{2}}\times 2^{\frac{D}{2}}$ arrays, that they can take their values in the two chiral representations,
repectively, of the algebra of $SO(D)$. It is therefore natural to introduce the full $SO(D)$ connection
\be
A_{i}=\left[
\begin{array}{cc}
A_{i}^{(+)} & 0\\
0 & A_{\mu}^{(-)}
\end{array}
\right]\,.\label{ASOD}
\ee
Next, we define the $D-$component $isovector$ Higgs field
\be
\Phi=\left[
\begin{array}{cc}
0 & \varphi\\
-\varphi^{\dagger} & 0
\end{array}
\right]=\f^i\,\Gamma_{i,D+1}\label{PHY}
\ee
in terms of the Dirac matrix representation of the algebra of $SO(D+1)$,
with $\Gamma_{i,D+1}=-\frac12\Gamma_{D+1}\Gamma_i$.

Note here the formal equivalence between the Higgs multiplet \re{PHY} in even $D$, to the corresponding one \re{Hisoodd}
in odd $D$. This formal equivalence turns out to be very useful in the calulus employed in following Sections.
In contrast with the former case of odd $D$ however, the form \re{PHY} for even $D$ is much more restrictive.
This is because in this case the Higgs multiplet is restricted to take its values in the components $\Gamma_{i,D+1}$
orthogonal to the elements $\Gamma_{ij}$ of $SO(D)$ by definition, irrespective of what symmetry is imposed.
It is clear that relaxing the spherical symmetry here, does not result in $\F$ getting out
of the orthogonal complement of $\Gamma_{ij}$, when $D$ is even.

From \re{ASOD}, follows the $SO(D)$ curvature
\be
F_{ij}=\pa_iA_j-\pa_jA_i+[A_i,A_j]=\left[
\begin{array}{cc}
F_{ij}^{(+)} & 0\\
0 & F_{ij}^{(-)}
\end{array}
\right]\label{FSOD}
\ee
and from \re{ASOD} and \re{PHY} follows the covariant derivative
\be
D_{\mu}\Phi=\pa_i\F+[A_i,\F]=\left[
\begin{array}{cc}
0 & D_{i}\varphi\\
-D_{i}\varphi^{\dagger} & 0
\end{array}
\right]\,.
\label{covPHY}
\ee
From \re{PHY} there simply follows the definition of $S$ for even $D$
\be
\label{SPHY}
S=-(\eta^2\,\eins+\F^2)=\left[
\begin{array}{cc}
S^{(+)} & 0 \\
0 & S^{(-)}
\end{array}
\right]\,.
\ee


\section{New Chern--Simons terms}
\label{CS}
First, we recall the usual dynamical Chern--Simons in odd dimensions defined in terms of the non-Abelian gauge connection.
Topologically massive gauge field theories in $2+1$ dimensional spacetimes were first introduced in
\cite{Deser:1982vy,Deser:1981wh}. The salient feature of these theories is the presence of a Chern-Simons (CS)
dynamical term. To define a CS density one needs to have a gauge connection, and hence also a curvature. Thus,
CS densities can be defined both for Abelian (Maxwell) and non-Abelian (Yang--Mills) fields. They can also be defined
for the gravitational~\cite{Jackiw:2003pm} field since in that system too one has a (Levi-Civita or otherwise) connection,
akin to the Yang-Mills connection in that it carries frame indices analogous to the isotopic indices of ther YM
connection. Here we are interested exclusively in the (non-Abelian) YM case, in the presence of an $isovector$ valued
Higgs field.

The definition of a Chern-Simons (CS) density follows from the definition of the corresponding Chern-Pontryagin (CP)
density \re{CP}. As stated by \re{totdivn}, this quantity is a total divergence and the density
$\bOmega^{(n)}=\Omega^{(n)}_M$ ($M=1,2,\dots,2n$)
in that case has $(2n)-$components. The Chern-Simons density is then defined as one fixed component of
$\bOmega^{(n)}$, say the $2n-$th component,
\be
\label{CS}
\Omega^{(n)}_{\rm{CS}}=\Omega_{2n}^{(n)}
\ee
which now is given in one dimension less, where $M=\mu,2n$ and $\mu=1,2,\dots (2n-1)$.

This definition of a (dynamical) CS term holds in all odd dimensional spacetimes
$(t,\R^D)$, with $x_{\mu}=(x_0,x_i)$, $i=1,2,\dots,D$, with $D$ being an even integer. That $D$ must be even is clear
since $D+2=2n$, the $2n$ dimensions in which the CP density \re{CP} is defined, is itself even.

The properties of CS densities are reviewed in \cite{Jackiw:1985}.
Most remarkably, CS densities are defined in odd (space or spacetime) dimensions and are $gauge$ $variant$. The context
here is that of a $(2n-1)-$dimensional Minkowskian space. It is important to realise that dynamical Chern-Simons theories are
defined on spacetimes with Minkowskian signature. The reason is that the usual CS densities appearing in the
Lagrangian are by construction $gauge$ $variant$, but in the definition of the energy densities the CS term itself does
not feature, resulting in a Hamiltonian (and hence energy) being $gauge$ $invariant$ as it should be~\footnote{Should one
employ a CS density on a space with Euclidean signature, with the CS density appearing in
the static Hamiltonian itself, then the energy would not be $gauge$ $invariant$. Hamiltonians of this
type have been considered in the literature, $e.g.$, in \cite{Rubakov:1986am}. Chern-Simons densities
on Euclidean spaces, defined in terms of the composite connection of a sigma model, find application as the
topological charge densities of Hopf solitons.}.

Of course, the CP densities and the resulting CS densities, can be defined in terms of both Abelian and non-Abelian
gauge connections and curvatures. The context of the present notes is the construction of soliton
solutions~\footnote{The term soliton solutions here is used rather loosely, implying only the construction of
regular and finite energy solutions, without insisting on topological stability in general.},
unlike in \cite{Deser:1982vy,Deser:1981wh}. Thus in any given dimension,
our choice of gauge group must be made with due regard to
regularity, and the models chosen must be consistent with the Derrick scaling requirement for the finiteness of
energy. Accordingly, in all but $2+1$ dimensions, our considerations are restricted to non-Abelian gauge fields.

Clearly, such constructions can be extended to all odd dimensional spacetimes systematically. We list
$\Omega_{\rm CS}$, defind by \re{CS}, for $D=2,4,6$, familiar densities
\bea
\Omega_{\rm CS}^{(2)}&=&\vep_{\la\mu\nu}\mbox{Tr}\,
A_{\la}\left[F_{\mu\nu}-\frac23A_{\mu}A_{\nu}\right]\label{CS3}\\
\Omega_{\rm CS}^{(3)}&=&\vep_{\la\mu\nu\rho\si}\mbox{Tr}\,
A_{\la}\left[F_{\mu\nu}F_{\rho\si}-F_{\mu\nu}A_{\rho}A_{\si}+
\frac25A_{\mu}A_{\nu}A_{\rho}A_{\si}\right]\label{CS5}
\\
\Omega_{\rm CS}^{(4)}&=&\vep_{\la\mu\nu\rho\si\tau\ka}
\mbox{Tr}\,A_{\la}\bigg[F_{\mu\nu}F_{\rho\si}F_{\tau\ka}
-\frac45F_{\mu\nu}F_{\rho\si}A_{\tau}A_{\ka}-\frac25
F_{\mu\nu}A_{\rho}F_{\si\tau}A_{\ka}\nonumber\\
&&\qquad\qquad\qquad\qquad\qquad\qquad
+\frac45F_{\mu\nu}A_{\rho}A_{\si}A_{\tau}A_{\ka}-\frac{8}{35}
A_{\mu}A_{\nu}A_{\rho}A_{\si}A_{\tau}A_{\ka}\bigg]\,.\label{CS7}
\eea

Concerning the choice of gauge groups, one notes that the
CS term in $D+1$ dimensions features the product of $D$ powers of the (algebra valued) gauge field/connection infront
of the Trace, which would vanish if the gauge group $is\ not\ larger\ than$ $SO(D)$. In that case, the YM connection would
describe only a 'magnetic' component, with the 'electric' component necessary for the the nonvanishing of the CS
density would be absent. As in \cite{Brihaye:2009cc}, the
most convenient choice is $SO(D+2)$. Since $D$ is always even, the representation of $SO(D+2)$ are the $chiral$
representation in terms of (Dirac) spin matrices. This completes the definition of the usual non-Abelian Chern-Simons
densities in $D+1$ spacetimes.

From \re{CS3}-\re{CS7}, it is clear that the CS density is $gauge$ $variant$. The Euler"=Lagrange equations of the
CS density is nonetheless $gauge$ $invariant$, such that for the exaples \re{CS3}-\re{CS7} the corresponding arbitarry
variations are
\bea
\delta_{A_{\la}}\Omega_{\rm CS}^{(2)}&=&\vep_{\la\mu\nu}F_{\mu\nu}\label{ELCS3}\\
\delta_{A_{\la}}\Omega_{\rm CS}^{(3)}&=&\vep_{\la\mu\nu\rho\si}F_{\mu\nu}F_{\rho\si}\label{ELCS5}\\
\delta_{A_{\la}}\Omega_{\rm CS}^{(4)}&=&\vep_{\la\mu\nu\rho\si\ka\eta}F_{\mu\nu}F_{\rho\si}F_{\ka\eta}\,.\label{ELCS7}
\eea
This, and other interesting properties of CS densities are given in \cite{Jackiw:1985}. A remarkable property of a CS
density is its transformation under the action of an element, $g$,
of the (non-Abelian) gauge group. We list these for the two examples \re{CS3}-\re{CS5},
\bea
\Omega_{\rm CS}^{(2)}\to\tilde\Omega_{\rm CS}^{(2)}&=&\Omega_{\rm CS}^{(2)}-
\frac23\vep_{\la\mu\nu}\mbox{Tr}\,\al_{\la}\al_{\mu}\al_{\nu}
-2\vep_{\la\mu\nu}\,\pa_{\la}\mbox{Tr}\,\al_{\mu}\,A_{\nu}\label{gaugeCS3}\\
\Omega_{\rm CS}^{(3)}\to\tilde\Omega_{\rm CS}^{(3)}&=&\Omega_{\rm CS}^{(3)}-\frac25\,\vep_{\la\mu\nu\rho\si}
\mbox{Tr}\,\al_{\la}\al_{\mu}\al_{\nu}\al_{\rho}\al_{\si}\nonumber\\
&&+
2\,\vep_{\la\mu\nu\rho\si}\,\pa_{\la}\mbox{Tr}\,\al_{\mu}\bigg[
A_{\nu}\left(F_{\rho\si}-\frac12A_{\rho}A_{\si}\right)+\left(F_{\rho\si}-\frac12A_{\rho}A_{\si}\right)A_{\nu}\nonumber\\
&&\qquad\qquad\qquad\qquad\qquad\qquad\qquad\qquad-\frac12\,A_{\nu}\,\al_{\rho}\,A_{\si}-\al_{\nu}\,\al_{\rho}\,A_{\si}
\bigg]\,,\label{gaugeCS5}
\eea
where $\al_{\mu}=\pa_{\mu}g\,g^{-1}$, as distinct from the algebra valued quantity $\beta_{\mu}=g^{-1}\,\pa_{\mu}g$ that
appears as the inhomogeneous term in the gauge transformation of the non-Abelian curvature (in our convention).

As seen from \re{gaugeCS3}-\re{gaugeCS5}, the gauge variation of $\Omega_{\rm CS}$ consists of a term which is
explicitly a total divergence, and, another term
\be
\label{om}
\omega^{(n)}\simeq\vep_{\mu_1\mu_2\dots\mu_{2n-1}}\mbox{Tr}\,\al_{\mu_1}\al_{\mu_2}\dots\al_{\mu_{2n-1}}\,,
\ee
which is {\it effectively total divergence}, and in a concrete group representation parametrisation becomes
{\it explicitly total divergence}. This can be seen by subjecting \re{om} to variations with respect to the
function $g$, and taking into account the Lagrange multiplier term resulting from the (unitarity) constraint
$g^{\dagger}\,g=g\,g^{\dagger}=\eins$.

The volume integaral of the CS density then transforms under a gauge transformation as follows.
Given the appropriate asymptotic decay of the connection (and hence also the curvature), the surface integrals in
\re{gaugeCS3}-\re{gaugeCS5} vanish. The only contribution to the gauge variation of the CS action/energy then comes from
the integral of the density \re{om}, which (in the case of Euclidean signature) for the appropriate
choice of gauge group yields an integer, up to the angular volume as a multiplicative factor.

All above stated properties of the Chern-Simons (CS) density hold irrespective of the signature of the space. Here, the
signature is taken to be Minkowskian, such that the CS density in the Lagrangian does not contribute to the energy density
directly. As a consequence the energy of the soliton is gauge invariant and does not suffer the gauge transformation
\re{gaugeCS3}-\re{gaugeCS5}. Should a CS density be part of a static Hamiltonian (on a space of Euclidean signature),
then the energy of the soliton would change by a multiple of an integer.

\subsection{New Chern--Simons terms in all dimensions}
The plan to introduce a completely new type of Chern-Simons term. The usual CS densities $\bOmega_{\rm{CS}}^{(n)}$,
\re{CS}, are defined with reference to the total divergence expression \re{totdivn} of the $n-$th
Chern-Pontryagin density \re{CP}, as the $2n-$th component $\Omega_{2n}^{(n)}$ of the density $\bOmega^{(n)}$.
Likewise, the new CS terms are defined with reference to the total divergence expression \re{totdiv} of the
dimensionally reduced $n-$th CP density, with the dimension $D$ of the residual space replaced
formally by $\bar{D}$
\be
\label{totdivbar}
{\cal C}_{\bar{D}}^{(n)}={\bf\nabla}\cdot{\bf\Omega}^{(n,\bar{D})}\,.
\ee
The densities ${\bf\Omega}^{(n,\bar{D})}$ can be read off from ${\bf\Omega}^{(n,D)}$ given in Section {\bf 5}, with
the formal replacement $D\to\bar{D}$.
The new CS term is now identified as the $\bar{D}-$th component of ${\bf\Omega}^{(n,\bar{D})}$. The final step
in this identification is to assign the value $\bar{D}=D+2$, where $D$ is the spacelike dimension of the $D+1$
dimensional Minkowski space, with the new Chern-Simons term defined as
\be
\label{CSnew}
\tilde\Omega_{\rm CS}^{(n,D+1)}\stackrel{def}=\Omega^{(n,D+2)}_{D+2}
\ee
The departure of the new CS densitiess from the usual CS densities is stark, and these
differ in several essential respects from the usual ones described in the previous subsection. The most important
new features in question are
\begin{itemize}
\item
The field content of the new CS systems includes Higgs fields in addition to the Yang-Mills fields,
as a consequence of the dimensional reduction of gauge fields described in Section {\bf 4}. It should be
emphasised that the appearance of the Higgs field here is due to the imposition of symmetries in the descent mechanism,
in contrast with its presence in the models~\cite{Hong:1990yh,Jackiw:1990aw,NavarroLerida:2009dm} supporting $2+1$
dimensional CS vortices, where the Higgs field was introduced by hand with the expedient of satisfying the Derrick
scaling requirement.
\item
The usual dynamical CS densities defined with reference to the $n-$th CP density live in $2n-1$ dimensional Minkowski
space, $i.e.$, only in odd dimensional spacetime. By contrast, the new CS densities defined with reference to the $n-$th
CP densities live in $D+1$ dimensional Minkowski space, for all $D$ subject to
\be
\label{subject}
2n-2\ge D\ge 2\,,
\ee
 $i.e.$, in both odd, as well as even dimensions. Indeed, in any given $D$ there is an infinite tower of new CS
densities characterised by the integer $n$ subject to \re{subject}. This is perhaps the most important feature of
the new CS densities.
\item
The smallest simple group consistent with the nonvanishing of the $usual$ CS density in $2n-1$ dimensional spacetime
is $SO(2n)$, with the gauge connection taking its values in the $chiral$ Dirac representation. By contrast,
the gauge groups of the new CS densities in $D+1$ dimensional spacetime are fixed by the
prescription of the dimensional descent from which they result.
As $per$  the prescription of descent described in Section {\bf 4}, the gauge group now will be $SO(D+2)$,
independently of the integer $n$, while the Higgs field takes its values in the orthogonal complement of $SO(D+2)$
in $SO(D+3)$. As such, it forms an iso-$(D+2)-$vector multiplet. 
\item
Certain properties of the new CS densities are remarkably different for $D$ even and $D$ odd.
\begin{itemize}
\item
Odd $D$: Unlike in the usual case \re{CS3}-\re{CS5}, the new CS terms are $gauge\ invariant$.
The gauge fields are $SO(D+2)$ and the Higgs are in $SO(D+3)$. $D$ being odd,
$D+3$ is even and hence the fields can be parametrised with respect to the $chiral$ (Dirac)
representations of $SO(D+3)$. An important consequence of this is the fact that now, both (electric) $A_0$
and (magnetic) $A_i$ fields lie in the same isotopic multiplets, in contrast to the $pseudo-$dyons described in the
previous section.
\item
Even  $D$: The new CS terms now consist of a $gauge\ variant$ part expressed only in terms of the gauge field, and a
$gauge\ invariant$ part expressed in terms of both gauge and Higgs fields. The leading, $gauge\ variant$, term differs
from the corresponding usual CS terms \re{CS3}-\re{CS5} only due to the presence of a (chiral) $\Ga_{D+3}$ matrix infront
of the Trace. The gauge and Higgs fields are again in $SO(D+2)$ and in $SO(D+3)$ respectively, but now, $D$ being even
$D+3$ is odd and hence the fields are parametrised with respect to the (chirally doubled up) full Dirac representations of
$SO(D+3)$. Hence the appearance of the chiral matrix infront of the Trace.
\end{itemize}
\end{itemize}
As in the usual CS models, the regular finite energy solutions of the new CS models are not topologically stable. These
solutions can be constructed numerically.

Before proceeding to display some typcal examples in the Subsection following, it is in order to make a small diversion
at this point to make a clarification. The new CS densities proposed are functionals of both the Yang--Mills, and, the
"isovector'' Higgs field. Thus, the systems to be described below are Chern-Simons--Yang-Mills-Higgs models in
a very specific sense, namely that the Higgs field is an intrinsic part of the new CS density. This is in contrast with
Yang-Mills--Higgs-Chern-Simons or Maxwell--Higgs-Chern-Simons models in $2+1$ dimensional spacetimes that
have appeared ubiquitously in the literature. It is important to emphasise that the latter are entirely different
from the systems introduced here, simply because the CS densities they employ are the $usual$ ones, namely \re{CS3} or
more often its Abelian~\footnote{There are, of course, Abelian CS densities in all odd spacetime dimensions but these do not
concern us here since in all $D+1$ dimensions with $D=2n\ge 4$, no regular solitons can be constructed.} version
\[
\Omega_{\rm U(1)}^{(2)}=\vep_{\la\mu\nu}\,A_{\la}F_{\mu\nu}\,,
\]
while the CS densities employed here are $not$ simply functionals of the gauge field, but also of the (specific)
Higgs field. To put this in perspective, let us comment on the
well known $Abelian$ CS-Higgs solitons in $2+1$ dimensions constructed in \cite{Hong:1990yh,Jackiw:1990aw}
support self-dual vortices, which happen to be unique inasfar as
they are also topologically stable. (Their non-Abelian counterparts~\cite{NavarroLerida:2009dm} are not endowed with
topological stability.) The presence of the Higgs field in \cite{Hong:1990yh,Jackiw:1990aw,NavarroLerida:2009dm}
enables the Derrick scaling requirement to be satisfied by virtue of the presence of the Higgs self-interaction
potential. In the Abelian case in addition, it results in the topological stability of the vortices.
If it were not for the topological stability, it would not be
necessary to have a Higgs field merely to satisfy the Derrick scaling requirement. That can be achieved instead, $e.g.$,
by introducing a negative cosmological constant and/or gravity, as was done
in the $4+1$ dimensional case studied in \cite{Brihaye:2009cc}. Thus, the involvement of the Higgs field in
conventional ($usual$) Chern-Simons theories is not the only option. The reason for emphasising the optional status of
the Higgs field in the usual $2+1$ dimensional Chern-Simons--Higgs models is, that in the new models proposed here
the Higgs field is intrinsic to the definition of the (new) Chern-Simons density itself.

\subsection{Examples}
As discussed above, the new dynamical Chern-Simons densities
\[
\tilde\Omega_{\rm CS}^{(n,D+1)}[A_{\mu},\F]
\]
are characterised by the dimensionality of the space $D$ and the integer $n$ specifying the
dimension $2n$ of the bulk space from which the relevant residual system is arrived at.

The case $n=2$ is empty, since according to \re{subject} the largest spacetime in which a new CS density can be
constructed is $2n-2$, $i.e.$, in $1+1$ dimensional Minkowsky space which we ignore.

The case $n=3$ is not empty, and affords two nontrivial examples. The largest spacetime $2n-2$, in which a
new CS density can be constructed in this case is $3+1$ and the next in $2+1$ Minkowski space. These, are, repectively,
\bea
\tilde\Omega_{\rm CS}^{(3,3+1)}&=&\vep_{\mu\nu\rho\si}\,\mbox{Tr}\ F_{\mu\nu}\,F_{\rho\si}\,\F
\label{3,3+1}\\
\tilde\Omega_{\rm CS}^{(3,2+1)}&=&
\vep_{\mu\nu\la}\,\mbox{Tr}\,\gamma_5\,\left[-2\eta^2A_{\la}\left(F_{\mu\nu}
-\frac23A_{\mu}A_{\nu}\right)+\left(\F\,D_{\la}\F-D_{\la}\F\,\F\right)\,F_{\mu\nu}
\right]\,.\label{3,2+1}
\eea

The case $n=4$ affords four nontrivial examples, those in $5+1$, $4+1$, $3+1$ and $2+1$ Minkowski space. These are,
repectively,
\bea
\tilde\Omega_{\rm CS}^{(4,5+1)}&=&\vep_{\mu\nu\rho\si\tau\la}\,\mbox{Tr}\ F_{\mu\nu}\,F_{\rho\si}\,F_{\tau\la}\,\F\label{4,5+1}\\
\tilde\Omega_{\rm CS}^{(4,4+1)}&=&\vep_{\mu\nu\rho\si\la}\,\mbox{Tr}\,\Gamma_7
\bigg[A_{\la}\left(F_{\mu\nu}F_{\rho\si}-F_{\mu\nu}A_\rho{}A_{\si}+\frac25A_{\mu}A_{\nu}A_{\rho}A_{\si}\right)\nonumber\\
&&\qquad\qquad\qquad\qquad+
D_{\la}\F\left(\F F_{\mu\nu}F_{\rho\si}+F_{\mu\nu}\F F_{mn}+F_{\mu\nu}F_{mn}\F\right)\bigg]\label{4,4+1}\\
\tilde\Omega_{\rm CS}^{(4,3+1)}&=&\vep_{\mu\nu\rho\si}\,\mbox{Tr}\bigg[
\F\left(\eta^2\,F_{\mu\nu}F_{\rho\si}+\frac29\,\F^2\,F_{\mu\nu}F_{\rho\si}+\frac19\,F_{\mu\nu}\F^2F_{\rho\si}\right)
\nonumber\\
&&\qquad\qquad\qquad\qquad-\frac29\left(\F D_{\mu}\F D_{\nu}\F-D_{\mu}\F\F D_{\nu}\F+D_{\mu}\F D_{\nu}\F\F\right)F_{\rho\si}\bigg]\label{4,3+1}\\
\tilde\Omega_{\rm CS}^{(4,2+1)}&=&\vep_{\mu\nu\la}\,\mbox{Tr}\,\Gamma_5\,\bigg\{6\eta^4\,A_{\la}\left(F_{\mu\nu}-\frac23\,A_{\mu}\,A_{\nu}
\right)\nonumber\\
&&\qquad\quad-6\,\eta^2\left(\F\,D_{\la}\F-D_{\la}\F\,\F\right)\,F_{{\mu\nu}}\nonumber\\
&&\qquad\quad+\left[\left(\F^2\,D_{\la}\F\,\F-\F\,D_{\la}\F\,\F^2\right)-2\left(\F^3\,D_{\la}\F-D_{\la}\F\,\F^3\right)\right]F_{{\mu\nu}}\,.
\bigg\}\label{4,2+1}
\eea
It is clear that in any $D+1$ dimensional spacetime an infinite tower of CS densities
$\tilde\Omega_{\rm CS}^{(n,D+1)}$ can be defined, for all positive integers $n$. Of these,
those in even dimensional spacetimes are gauge invariant, $e.g.$, \re{3,3+1}, \re{4,5+1} and  \re{4,3+1},
while those in odd dimensional spacetimes are gauge variant, $e.g.$, \re{3,2+1}, \re{4,4+1} and \re{4,2+1}, the gauge variations in these cases being given formally by \re{gaugeCS3}
and \re{gaugeCS5},
with $g$ replaced by the appropriate gauge group here.

Static soliton solutions to models whose Lagrangians consist of the above introduced types of
CS terms together with Yang-Mills--Higgs (YMH) terms are currently under construction.
The only constraint in the choice of the detailed models employed is the requirement that the
Derrick scaling requirement be satisfied. Such solutions are constructed numerically. In
contrast to the monopole solutions, they are not endowed with topological stability
because the gauge group must be larger than $SO(D)$, for which the solutions to the
constituent YMH model is a stable monopole. Otherwise the CS term would vanish.
\\
\\
\noindent
{\bf\large Acknowledgement}

\noindent
This work is carried out in the framework of Science Foundation Ireland (SFI) project
RFP07-330PHY.

\begin{small}
 
\end{small}


\begin{thebibliography}{99}
\bibitem{D}
P.A.M. Dirac, Proc. Roy. Soc. A {\bf 133} (1931) 60.
\bibitem{yang}
C.~N.~Yang, J. Math. Phys. {\bf 19} (1978) 320.
\bibitem{Tchrakian:2008zz}
  T.~Tchrakian,
  Phys.\ Atom.\ Nucl.\  {\bf 71} (2008) 1116.
\bibitem{Schwarz:1977ix}
  A.~S.~Schwarz,
  Commun.\ Math.\ Phys.\  {\bf 56} (1977) 79.
\bibitem{Romanov:1977rr}
  V.~N.~Romanov, A.~S.~Schwarz and Yu.~S.~Tyupkin,
  Nucl.\ Phys.\  B {\bf 130} (1977) 209.
\bibitem{Schwarz:1981mb}
  A.~S.~Schwarz and Yu.~S.~Tyupkin,
  Nucl.\ Phys.\  B {\bf 187} (1981) 321.
\bibitem{Deser:1982vy}
  S.~Deser, R.~Jackiw and S.~Templeton,
  Phys.\ Rev.\ Lett.\  {\bf 48} (1982) 975.
\bibitem{Deser:1981wh}
  S.~Deser, R.~Jackiw and S.~Templeton,
  Annals Phys.\  {\bf 140} (1982) 372
  [Erratum-ibid.\  {\bf 185} (1988) 406]
  [Annals Phys.\  {\bf 185} (1988) 406]
  [Annals Phys.\  {\bf 281} (2000) 409].
\bibitem{Jackiw:2003pm}
  R.~Jackiw and S.~Y.~Pi,
  Phys.\ Rev.\  D {\bf 68} (2003) 104012
  [arXiv:gr-qc/0308071].
\bibitem{Jackiw:1985}
R. Jackiw, "Chern-Simons terms and cocycles in physics and mathematics", in E.S. Fradkin $Festschrift$, Adam Hilger, Bristol (1985)
\bibitem{Rubakov:1986am}
  V.~A.~Rubakov and A.~N.~Tavkhelidze,
  Phys.\ Lett.\  B {\bf 165} (1985) 109.
\bibitem{Brihaye:2009cc}
  Y.~Brihaye, E.~Radu and D.~H.~Tchrakian,
  Phys.\ Rev.\  D {\bf 81} (2010) 064005
  [arXiv:0911.0153 [hep-th]].
\bibitem{Hong:1990yh}
  J.~Hong, Y.~Kim and P.~Y.~Pac,
  Phys.\ Rev.\ Lett.\  {\bf 64} (1990) 2230.
\bibitem{Jackiw:1990aw}
  R.~Jackiw and E.~J.~Weinberg,
  Phys.\ Rev.\ Lett.\  {\bf 64} (1990) 2234.
\bibitem{NavarroLerida:2009dm}
  F.~Navarro-Lerida and D.~H.~Tchrakian,
  Phys.\ Rev.\  D {\bf 81} (2010) 127702
  [arXiv:0909.4220 [hep-th]].
  
  
  
  


\end{thebibliography}
\end{document}